\documentclass[twocolumn,showpacs,superscriptaddress]{revtex4}
\usepackage{graphicx}

\newcommand{\beq}{\begin{equation}}
\newcommand{\eeq}{\end{equation}}

\begin{document}



\title{Assessment of petrophysical quantities inspired by joint multifractal approach}

\author{Z. Koohi Lai }
\affiliation{Department of Physics, Islamic Azad University,
Firoozkooh Branch, Firoozkooh, Iran}

\author{ S. M. S. Movahed }
\affiliation{Department of Physics, Shahid Beheshti University, G.C., Evin, Tehran 19839, Iran
}

\author{ G. R. Jafari  }
\affiliation{Department of Physics, Shahid Beheshti University, G.C., Evin, Tehran 19839, Iran
}

\vskip 1cm

\begin{abstract}

In this paper joint multifractal random walk approach is carried out to analyze
some petrophysical quantities for characterizing the petroleum
reservoir. These quantities include Gamma emission (GR), sonic
transient time (DT) and Neutron porosity (NPHI) which are collected
from four wells of a reservoir. 
To quantify mutual interaction of petrophysical quantities, joint
multifractal random walk is implemented. This approach is based on
the mutual multiplicative cascade notion in the multifractal
formalism and in this approach $L_0$ represents a benchmark to
describe the nature of cross-correlation between two series. The
analysis of the petrophysical quantities revealed that GR for all
wells has strongly multifractal nature due to the considerable
abundance of large fluctuations in various scales.  The variance of probability distribution function,
$\lambda_{\ell}^2$, at scale $\ell$ and its intercept determine the
multifractal properties of the data sets sourced by probability density function. The value of
$\lambda_0 ^2$ for NPHI data set is less than GR's, however, DT
shows a nearly monofractal behavior, namely $\lambda_0 ^2\rightarrow
0$, so we find that $\lambda_0^2({\rm GR})>\lambda_0^2({\rm
NPHI})\gg\lambda_0^2({\rm DT})$. While, the value of
Hurst exponents can not discriminate between series GR, NPHI and DT.
Joint analysis of the petrophysical quantities for considered wells
demonstrates that $L_0$ has negative value for GR-NPHI confirming
that finding shaly layers is in competition  with finding porous
medium while it takes positive value for GR-DT determining that
continuum medium can be detectable by evaluating the statistical
properties of GR and its cross-correlation to DT signal.\\



\end{abstract}
\maketitle

\section{Introduction}
Undoubtedly, petroleum, gas and  fossil fuels have most important
impact on economics, social life and associated industries
\cite{laz12}. In the research of oil and gas fields, petrophysical
quantities are analyzed in order to determine the economic benefit
of oil fields and gas production and consequently on decision what
equipments are useful to improve the extract and/or production
efficiency of underlying wells. For a typical reservoir the
characteristics such as thickness (bed boundaries), lithology
containing information about rock type, porosity, fluid saturations,
fluid identification and permeability, pressure and fractional flow
involving gas, oil and water should be quantify as accurately as
possible. There are several indicators to explore and analyze oil
and gas reservoirs \cite{ellis87,datalog99,hea00,rashed11}. We are
not able to extract full information from them without understanding
how they are affected by each others. These indicators possess a
non-Gaussian behavior due to the fact that the density of oil wells
depend on depth of reservoirs. By getting close to oil reservoir a
gradient in the medium is observed. This kind of non-Gaussianity
could be a sign of medium changes and/or an indicator of reservoir
approaching. Prospect benchmarks in such system are not only coupled
but also may be non-Gaussian. In order to take into account both
mentioned  properties in a typical system, simultaneously, the
generalized multifractal random walk can be a proper method to
implement \cite{Bacry,Muzy1,Bacry1,Muzy2}.

The multifractal formalism introduced in the theory of complex
systems and nonlinear dynamics has been applied in various fields of
researches ranging from biology and finance e.g. foreign exchange
rates \cite{Ghas96}, stock index \cite{Kiy06,Sadegh}, human
heartbeat fluctuations \cite{Kiy004}, seismic time series
\cite{Tabar09,Telesca0,Telesca1}, sol-gel transition \cite{Shayeganfar2},
non-equilibrium growth processes \cite{vicsek,barabashi} and solar
and wind energies \cite{Sorrise,Telesca} to climate and metrology
\cite{climate1,climate2,climate3,climate4,sadeghriver}. However the notion of multifractality, is widely
used in many of above researches, but there are different approaches
to characterize this concept in such systems from complexity point
of view.

Multifractal models have been developed inspired by turbulent
cascades in hydrodynamic turbulence in which multiplicative cascades
display scale-invariant statistical properties \cite{Frisch}. In the
context of multiplicative random cascades
\cite{Meneveau1,halsey,palad,Meneveau}, recently Bacry et al.
introduced multifractal random walk  model as a continuous
random walk with the logarithm of the correlated stochastic
variances \cite{Bacry,Muzy1,Bacry1}. The occurrence of large fluctuations
in a typical system leads to a log-normal deviation from the normal
shape of probability density function. Consequently, multifractality is imposed in such
system sourced by deviation from Gaussian PDF. The mutual interaction between various fluctuations in
linear and non-linear regimes are of interest, so in order to
examine such property, Muzy et al. \cite{Muzy2} demonstrated a
generalization of univariate multifractal random walk to a multivariate framework which
is so-called multivariate multifractal random walk.

In this paper we follow the research done by Z. Koohi et. al.
\cite{koohi}, and rely on joint multifractal random walk approach to make more complete
our knowledge concerning petrophysical data sets. In \cite{koohi},
authors examined the shape of probability distribution function
(PDF) of underlying quantity in the framework of multiplicative
random cascades. Also the changing shape of PDF with chosen scale,
namely from dissipation to large scales, has been characterized by
finding the scale dependency of $\lambda^2$ denoted by
$\lambda_{\ell} ^2$. The so-called non-Gaussian factor,
$\lambda_{\ell} ^2$, characterizes the non-Gaussianity of
corresponding PDF. In turbulence this quantity deals with number of
Cascades \cite{naert1,naert2}. From thermodynamics point of view,
$\lambda_{\ell} ^2$ is potentially related to partition function,
therefore free energy of system can be pertinent as:
$F=-k_{\beta}T\ln(\lambda_{\ell} ^2)$ \cite{novikov}. In addition,
parameter $\lambda_{\ell} ^2$ represents fluctuations of the
variances based on the notion of log-normal multiplicative
processes. This means that, the larger $\lambda_{\ell} ^2$, the
higher probability of finding higher values of fluctuation in
underlying quantity. Besides mentioned investigations, the mutual
correlation between various petrophysical surveys have been motivated from statistical properties point of view. To this end,
joint multifractal random walk will be implemented to examine cross-correlation
properties.

The quantities investigated in this research include Gamma emission,
(GR), sonic transient time, (DT), and Neutron porosity, (NPHI)
\cite{ellis87,datalog99,hea00,Leite,Montagne,Christie,koohi12,Fedi,log1,log2}.
Each of mentioned data contains valuable information about the
underlying reservoir. GR is capable to give proper estimation
concerning radioactive components existing in reservoir rocks. NPHI
belongs to category of nuclear logging providing some information
about porosity and lithology and has been established on migration
length and bulk capture cross-section. Our results obtained from
four wells of the reservoir confirmed that GR has strongly
multifractal behavior due to the considerable abundance of large
scale fluctuations in this quantity. The value of $\lambda_0 ^2$
($\lambda_0 ^2$ is intercept of $\lambda_{\ell}^2$ versus $\ell$)
for NPHI data set is less than GR's, however, DT shows a nearly
monofractal behavior ($\lambda_0 ^2\rightarrow 0$). From joint
analysis point of view, our results represent that $L_0=L_{\ell\to
0}$ for GR-NPHI pair has negative value while a positive value of
$L_0$ for GR-DT pair is attained. To make theoretical prediction for
scaling exponents computed by multifractal random walk and its joint
method, we also apply adaptive detrending method to remove probable
trends superimposed on data and clean date will be used for
Detrended Fluctuation Analysis (DFA) to determine corresponding
Hurst exponent. This exponent is used for deriving theoretical
prediction of scaling exponents derived in context of multifractal
random walk.

The rest of this paper is organized as follows: In sections
\ref{model1} and \ref{model2}, we describe multifractal random walk
model and joint multifractal random walk, respectively. In section
\ref{adap111}, Adaptive detrending algorithm followed by Detrending
Fluctuation Analysis are explained. In section \ref{data1} we
explain the data sets and the location where they have been
recorded. Section \ref{data} is devoted to analyze the petrophysical
data sets. Summary and conclusion are given in section \ref{sum}.

\section{Multifractal model}\label{model1}

In this section we rely on multifractal approach to model the
underlying data set. Self-similarity and self-affinity can be
assigned to many observed shapes as well as processes in nature.
This geometrical index was introduced for the first time by
Mandelbrot \cite{Feder}. The particular characteristic concerning
fractal and multifractal phenomena is scaling behavior. Assume a
typical stochastic fluctuation recorded during an experiment or
simulation as a function of dynamical parameter (spatial or
temporal) represented by $x(t)$. One of scale invariant properties
of mentioned stochastic series is generally demanded by $\xi_q$ as
follows \cite{Bacry}:
\begin{eqnarray}\label{eq1}
m(q,\ell)&\equiv&\sum_{t} \mid x(t+\ell)-x(t)\mid ^{q} \cr \nonumber \\
&=&\mathcal{A}_{q}\ell^{\xi_{q}}
\end{eqnarray}
here $\mathcal{A}_q$ is a prefactor and $\xi_{q}$ corresponds to the
exponent of power law function. If the exponent $\xi_{q}$ is a linear
function versus $q$, namely $\xi_{q} = Hq$, a single Hurst exponent,
$H$, is adequate to characterize the fractal property of underlying
signal and $x(t)$ is called monofractal. While for a nonlinear
dependence of $\xi_{q}$ with respect to $q$, $x(t)$ belongs to
multifractal category. It must point out that the range of scaling
regime might be given as a prior, namely for turbulence, it is less
than characteristic scale in fully developed turbulence. For
arbitrary fluctuation, mentioned length (time) scale is considered
about correlation length (time) scale. According to physics of
turbulence, it has been demonstrated that for small Reynolds number,
the inertial range is very small and the scaling behavior described
by Eq. (\ref{eq1}) is either absent or difficult to observe
\cite{Gao}. Therefore, in a general process, this case may occur.
The concept of extended self-similarity provides a solution to this
problem. Benzi et al. found that the scaling properties of the
velocity increments can be extended up to the dissipation range if
we modify Eq. (\ref{eq1}) as : $m(q,\ell)\sim m(3,\ell)^{\zeta_{q}}$
\cite{Benzi}. For fractal processes  $m(3,\ell)\sim \ell^{3H}$ then
$\zeta_{q}=\frac {q} {3}$. The relation between $\xi_{q}$ as well as
$\zeta_{q}$ and the non-Gaussian parameter in the hierarchical
multiplicative cascade model developed for the first time by
Castaing et al. \cite{Castaing}. In this robust approach the
multifractality is assigned to PDF of underlying data set
\cite{frisch97,Mandelbrot,Kahane,Meneveau,Hentschel}. Suppose that
the increment of fluctuations at scales $\ell$ and $\beta \times
\ell\ (\beta <\ell)$ can be modeled through the cascading rule:
\begin{equation}\label{eq3}
 [x(t+\beta\times \ell)-x(t)]=W_{\beta}  [x(t+\ell)-x(t)], \quad \forall \ \ell,\beta >0
\end{equation}
here $W_{\beta}$ is a stochastic variable
depending only on $\beta$ and, behaves as a logarithmic infinitely
divisible law \cite{Castaing,Arneodo2}. Hereafter, for convenient,
we use, $\delta_{\beta \times \ell}x(t)\equiv [x(t+\beta \times
\ell)-x(t)]$ and $\delta_{\ell}x(t)\equiv [x(t+\ell)-x(t)]$. In
addition, the integral form of the corresponding PDF at scale $\ell$
using its increment at scale $\beta \times \ell (\beta<\ell)$ can be
written as \cite{Castaing}:
\begin{equation}\label{eq4}
P_{\ell} (\delta_{\ell} x)=\int G_{\ell,\beta \times \ell} (u) e^{-u}
P_{\beta \times \ell} (e^{-u} \delta_{\ell}x) du
\end{equation}
This equation states that PDF of $\delta_{\ell}x$ at a given scale,
$\ell$, is determined as a weighted sum of PDF at a larger scale,
$\beta \times \ell$. The shape of weight function, $G_{\ell,\beta
\times \ell} (u)$, is determined by statistical nature of underlying
data. As an example, for a self-similar kernel with a given Hurst
exponent, the shape of kernel reads as $G_{\ell,\beta \times \ell}
(u)=\delta_{D}(u-H\ln(\ell/(\beta \times \ell)))$. Here $\delta_D$
is Dirac delta function. Subsequently, $P_{\ell} (\delta_{\ell}
x)\sim\beta^H P_{\beta \times \ell}(\beta^H \delta_{\ell}x)$ which
is known as a geometrical convolution between the kernel
$G_{\ell,\beta \times \ell}$ and $P_{\beta \times \ell}$
\cite{Muzy1}. Eq. (\ref{eq4}) enables us to calculate $q$th order of
absolute moment, $m(q,\ell)$ by determining the functional form of
kernel. Any deviation from Dirac delta function for kernel leads to
a deviation from pure Gaussian function for $P_{\ell} (\delta_{\ell}
x)$. In this case underlying data has multifractal nature,
consequently, the corresponding $\xi_q$ deviates from the linear
behavior versus $q$. Inspired by fully developed turbulent flows
by Castaing et. al. \cite{Castaing}, one can find  various
stochastic variables which their PDFs are the same as that of given
by Eq. (\ref{eq4}). As an example one can notice to
\cite{Castaing,frisch97,Bacry}:
\begin{equation}
\delta _{\ell}x(t)=\mathcal{B}_{\ell}(t) e^{\omega _{\ell}(t)}
\end{equation}
The PDFs of $\mathcal{B}_{\ell}(t)$ and $\omega _{\ell}(t)$ are
Gaussian and the mean value of both variables is zero. The variances
of stochastic variables, $\mathcal{B}_{\ell}(t)$ and $\omega
_{\ell}(t)$ are $\sigma_{\ell} ^2$ and $\lambda_{\ell} ^2$,
respectively. Therefore PDF of mentioned stochastic variable becomes
\cite{Castaing}:
\begin{equation}\label{eq6}
P_{\ell} (\delta_{\ell} x)=\int _{0} ^{\infty}G_{\ell}(\ln
\sigma_{\ell}) \frac{1} {\sigma_{\ell}} F_{\ell}\left(\frac {\delta
_{\ell}x} {\sigma_{\ell}}\right) d(\ln \sigma_{\ell})
\end{equation}
where
\begin{eqnarray}\label{gpdf}
G_{\ell}(\ln \sigma_{\ell})&=&\frac {1} {\sqrt{2\pi} \lambda
_{\ell}} \exp\left(-\frac {\ln ^{2}\sigma_{\ell}} {2\lambda _{\ell}
^{2}}\right)
\end{eqnarray}
\begin{eqnarray}\label{fpdf}
F_{\ell} \left(\frac {\delta _{\ell}x}
{\sigma_{\ell}}\right)&=&\frac {1} {\sqrt{2\pi}} \exp\left(-\frac
{\delta_{\ell} x^{2}} {2\sigma_{\ell} ^{2}}\right)
\end{eqnarray}
$P_{\ell} (\delta_{\ell} x)$ converges to a Gaussian function when
$\lambda_{\ell}\rightarrow 0$. The expectation value of various
order of increment reads as:
\begin{eqnarray}\label{mqq}
m(q,\ell)&=&\int  |x(t+\ell)-x(t)|^q P_{\ell}(\delta_{\ell} x)d(\delta_{\ell} x)
\end{eqnarray}
Using Eqs. (\ref{eq6}) and (\ref{mqq}) the scaling exponent defined
in Eq. (\ref{eq1}) is given by \cite{Bacry}:
\begin{equation}\label{eq8}
\xi _{q}=qH-q(q-1)\frac{\lambda _{0} ^{2}} {2}
\end{equation}
where $\lambda _{0} ^{2}$ is determined by intercept of
$\lambda _{\ell} ^{2}$ as a function of $\ell$ \cite{Kiyono2,F}.
The prefactor in Eq. (\ref{eq1}) is also calculated by:
\begin{equation}
\mathcal{A}_{q}=\int_{-\infty} ^{+\infty} x^q F(x)dx
\end{equation}
here $F(:)$ is indicated by Eq. (\ref{fpdf}). In general case,
according to the multiplicative cascading processes starting from
large scale, $\mathcal{L}$, to small scale by supposing the scaling relation
$\beta = \frac {1} {2} $, $m(q,\ell)$ holds for all range $\ell_{n}
=\beta ^{n} \mathcal{L}$ \cite{Bacry}.

Also the correlation function of various order of increment at
scale $\ell$ in terms of length (time) lag $\tau$ is:
\begin{equation}\label{eq10}
C_{\ell} ^{q} (\tau )\equiv \langle | x(t+\ell)-x(t)|^{q}
|x(t+\ell+\tau )-x(t)|^{q}\rangle
\end{equation}
with $\ell<\tau$ then by using Eqs. (\ref{eq6}), (\ref{gpdf}) and (\ref{fpdf}), Eq. (\ref{eq10}) becomes \cite{Bacry}:
\begin{equation}
C_{\ell} ^{q} (\tau )\sim \mathcal{A}_q\left(\frac {\tau}
{L}\right)^{2\xi _{q}} \left(\frac {\ell} {L}\right)^{-q^{2} \lambda
_{0} ^{2}}
\end{equation}

In the next section, we will
explain the modified version of multifractal random walk in the context of cross-correlation, namely joint multifractal random walk.\\

\section{Joint Multifractal Random Walk}\label{model2}
There are many approaches to investigate the mutual effect of two
processes, such as Detrended Cross-Correlation Analysis \cite{DCCA}
and its generalized Multifractal Detrended Cross-Correlation
Analysis \cite{mf-dxa}. Here we rely on multifractal random walk
approach generalized by Muzy et al. \cite{Muzy2}. This
generalization takes into account the cross-correlations of
stochastic variances for two processes. Suppose that
$\textbf{x}=\{x_{1}(t),x_{2}(t)\}$ is a bivariate process, with
regard to  cascading rule (Eq. (\ref{eq3})), one can write the
bivariate cascading relation by \cite{Bacry1,Muzy2}:
\begin{equation}
\delta_{\beta \times \ell} x_{i} (t)=W_{\beta,i} \delta_{\ell}
x_{i}(t) \ \ \forall \ \ell,\beta >0 \ \ {\rm{and}} \ \ i=\{1,2\}
\end{equation}
here $\textbf{W}\equiv\{W_{\beta,1},W_{\beta,2}\}$ is a log
infinitely divisible stochastic vector which depends only on
$\beta$. The bivariate version of multifractal random walk is
defined as \cite{Bacry1,Muzy2}:

\begin{equation}
\delta_{\ell}\textbf{x}(t)=\delta_{\ell}x_{1}(t)\times
\delta_{\ell}x_{2}(t)=\left(\mathcal{B}_{1}^{(\ell)}(t)e^{\omega_{1}^{(\ell)}(t)},\mathcal{B}_{2}^{(\ell)}(t)e^{\omega_{2}^{(\ell)}(t)}\right)
\end{equation}
where $(\mathcal{B}_{1}^{(\ell)},\mathcal{B}_{2}^{(\ell)})$ and
$(\omega_{1}^{(\ell)},\omega_{2}^{(\ell)})$ have both joint Gaussian
probability density function with zero mean. The covariance matrix
of  $(\mathcal{B}_{1}^{(\ell)},\mathcal{B}_{2}^{(\ell)})$ is
$\mathbf{\Sigma}_{(\ell)}$ which is defined according to:
\begin{equation}
\mathbf{\Sigma}_{(\ell)}\equiv\left(
               \begin{array}{cc}
                 \Sigma_{(\ell)}^{11}& \Sigma_{(\ell)}^{12}\\
                 \Sigma_{(\ell)}^{21} & \Sigma_{(\ell)}^{22} \\
               \end{array}
             \right)
\end{equation}
This is so-called Markowitz matrix \cite{Muzy2} and
$\mathbf{\Lambda}_{(\ell)}$ represents the covariance matrix of
$(\omega_{1}^{(\ell)},\omega_{2}^{(\ell)})$ indicated as:
\begin{equation}
             \mathbf{\Lambda}_{(\ell)}\equiv\left(
                                                         \begin{array}{cc}
                                                           \Lambda_{(\ell)}^{11} & \Lambda_{(\ell)}^{12} \\
                                                           \Lambda_{(\ell)}^{21} & \Lambda_{(\ell)}^{22} \\
                                                         \end{array}
                                                       \right)
\end{equation}
where $\Sigma_{(\ell)}^{11}\equiv\sigma_1^2(\ell),
\Sigma_{(\ell)}^{22}\equiv\sigma_2^2(\ell)$ and
$\Lambda_{(\ell)}^{11}\equiv\lambda_1 ^2(\ell) ,
\Lambda_{(\ell)}^{22}\equiv\lambda_2 ^2(\ell)$.  In addition, the
off-diagonal terms $\Lambda_{(\ell)}^{12}$ and
$\Sigma_{(\ell)}^{12}$ satisfy the relations
$\Lambda_{(\ell)}^{12}=\Lambda_{(\ell)}^{21}=L_{\ell}\lambda_1(\ell)\lambda_2(\ell)$
and $\Sigma_{(\ell)}^{12}=\Sigma_{(\ell)}^{21}=S_{\ell}
\sigma_1(\ell)\sigma_2(\ell)$. The above matrix is known as
Multifractal matrix \cite{Muzy2}. The PDFs of
$(\mathcal{B}_{1}^{(\ell)},\mathcal{B}_{2}^{(\ell)})$ and
$(\omega_{1}^{(\ell)},\omega_{2}^{(\ell)})$ have the following form:
\begin{eqnarray}\label{eq15}
F_{\ell}(\mathcal{B}_{1}^{(\ell)},\mathcal{B}_{2}^{(\ell)})&=&\frac{1}{2\pi\sqrt{\rm{Det}(\mathbf{\Sigma}_{(\ell)})}}\
\exp\left({-\frac{\mathcal{B}_{(\ell)}^T \cdot\mathbf{\Sigma}_{(\ell)}^{-1}\cdot \mathcal{B}_{(\ell)}}{2}}\right) \nonumber\\
\end{eqnarray}
\begin{eqnarray}\label{eq151}
G_{\ell}(\omega_{1}^{(\ell)},\omega_{2}^{(\ell)})&=&\frac{1}{2\pi\sqrt{\rm{Det}(\mathbf{\Lambda}_{(\ell)})}}\
\exp\left({-\frac{\mathbf{\omega}^T_{(\ell)}\cdot\mathbf{\Lambda}^{-1}_{(\ell)}\cdot\mathbf{\omega}_{(\ell)}}{2}}\right)\nonumber\\
\end{eqnarray}
Therefore the joint PDF of stochastic vector is:
\begin{widetext}
\begin{equation}\label{eq16}
P_\ell(\delta_{\ell} x_{1},\delta_{\ell} x_{2})=\int d(\ln
\sigma_{1}(\ell)) \int d(\ln \sigma_{2}(\ell))G_{\ell}\left(\ln
\sigma_{1}(\ell) , \ln \sigma_{2}(\ell)\right) \frac{1}
{\sigma_{1}(\ell)\sigma_{2}(\ell)} F\left(\frac {\delta_{\ell}
x_{1}} {\sigma_{1}(\ell)},\frac {\delta_{\ell} x_{2}}
{\sigma_{2}(\ell)}\right)
\end{equation}
\end{widetext}

According to Eqs. (\ref{eq15}) and (\ref{eq151}), one can write
$G_{\ell}(\ln \sigma_{1}(\ell) , \ln \sigma_{2}(\ell))$ and
$F_{\ell}(\frac {\delta_{\ell} x_{1}} {\sigma_{1}(\ell)},\frac
{\delta_{\ell} x_{2}} {\sigma_{2}(\ell)})$ as:
\begin{widetext}
\begin{eqnarray}
G_{\ell}\left(\ln \sigma_{1}(\ell) , \ln
\sigma_{2}(\ell)\right)=\frac {1} {2\pi\lambda _{1}(\ell) \lambda
_{2}(\ell) \sqrt{(1-L_{\ell} ^{2})}} \exp\left(-\frac {1}
{2(1-L_{\ell} ^{2})} \left[ \left(\frac {\ln ^{2}\sigma_{1}(\ell)}
{\lambda _{1} ^{2}(\ell)}\right)+\left(\frac {\ln
^{2}\sigma_{2}(\ell)} {\lambda _{2} ^{2}(\ell)}\right)-
\frac{2L_{\ell} (\ln{\sigma_{1}(\ell)}\ln{\sigma_{2}(\ell)})}
{\lambda _{1}(\ell)
\lambda _{2}(\ell)} \right]\right)\nonumber\\
\end{eqnarray}
\begin{eqnarray}\label{fjoint}
F_{\ell}\left(\frac {\delta_{\ell} x_{1}} {\sigma_{1}(\ell)},\frac
{\delta_{\ell} x_{2}} {\sigma_{2}(\ell)}\right)=\frac {1} {2\pi
\sqrt{(1-S_{\ell} ^{2})}} \exp\left(-\frac {1} {2(1-S_{\ell} ^{2})}
\left[ \left(\frac {(\delta_{\ell} x_{1})^{2}} {\sigma_{1}
^{2}(\ell)}\right)+\left(\frac {(\delta_{\ell} x_{2})^{2}}
{\sigma_{2} ^{2}(\ell)}\right)- 2S_{\ell} \left(\frac{\delta_{\ell}
x_{1}} {\sigma_{1}(\ell)}\right)\left(\frac{\delta_{\ell} x_{2}}
{\sigma_{2}(\ell)}\right) \right]\right)\nonumber\\
\end{eqnarray}
\end{widetext}
$P_{\ell}(\delta_{\ell} x_{1},\delta_{\ell} x_{2})$ takes the product of
$P_{\ell}(\delta_\ell x_{1})P_{\ell}(\delta_\ell x_{2})$ when
$L_{\ell}$ and $S_{\ell}$ tend to zero. By demanding the scale invariant feature for
the joint moment of two processes $\delta_{\ell}x_{1}$ and
$\delta_{\ell}x_{2}$ according to \cite{Muzy2}:

\begin{eqnarray}\label{xijoint1}
m_{\rm{joint}}(q_{1},q_{2};\ell)&=&\langle |\delta_{\ell}x_{1}|^{q_{1}}
|\delta_{\ell}x_{2}|^{q_{2}} \rangle\nonumber\\
 &=& \mathcal{A}_{q_{1},q_{2}}\
\ell^{\xi^{\rm{joint}} _{q_{1}q_{2}}}
\end{eqnarray}
one can show that the scaling exponent $\xi^{\rm{joint}} _{q_{1}q_{2}}$ and
prefactor $\mathcal{A}_{q_{1},q_{2}}$ are given by (see appendix for more
details):
\begin{equation}\label{eq19}
\xi^{\rm{joint}}_{q_{1}q_{2}}=\xi_{q_{1}}^{(1)}+\xi_{q_{2}}^{(2)}-L_{0}
q_{1} q_{2}
\end{equation}
and
\begin{equation}\label{eq20}
\mathcal{A}_{q_{1},q_{2}}=\int_{-\infty} ^{+\infty}\int_{-\infty} ^{+\infty}
x^{q_1} x'^{q_2} F(x,x')dx dx'
\end{equation}
where $F(x,x')$ is given by Eq. (\ref{fjoint}). The exponents
$\xi_{q_{1}}^{(1)}$ and $\xi_{q_{2}}^{(2)}$ refer to the scaling
exponents of the first and second processes, respectively and they
are determined via Eq. (\ref{eq8}). Parameter $L_0$ is determined by
intercept of $L_{\ell}$ as a function of ${\ell}$. If two processes
are independent then $L_0\rightarrow 0$, consequently,
$\xi^{\rm{joint}}_{q_{1}q_{2}}=\xi_{q_{1}}^{(1)}+\xi_{q_{2}}^{(2)}$.

\section{Adaptive Detrended Fluctuation Analysis}\label{adap111}

\begin{figure}
\begin{center}
\includegraphics[width=0.9\linewidth]{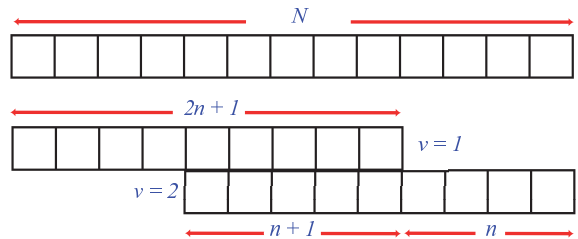}
\caption{\label{adap} A schematic to clarify how the adaptive detrending method is implemented  on desired series.}
\end{center}
\end{figure}

It turns out that data sets recording in the nature are affected by
trends and unknown noises. To compute reliable physical quantities,
not only we should improve the quality of tools in order to reduce
systematic errors, but also robust methods in data analysis
containing high performance algorithm and capable to exclude the
undesired trends and noises should necessary. To this end, Detrended
Fluctuation Analysis (DFA) was introduced \cite{peng92,peng94}. But
unfortunately, the effect of various kinds of trends on scaling
behavior of fluctuation function remains debatable
\cite{hu09,zhou10,hui12}. Here to resolve this discrepancy as much
as possible, we apply adaptive detrended algorithm as a
complementary producer to extract the superimposed trend on
underlying data sets. Since adaptive detrending method and DFA are
used as a complementary algorithms so hereafter we call them as
Adaptive-DFA method. The Adaptive-DFA method contains five steps
(see \cite{peng92,peng94,hu09,bul95,bunde02}
for more details):\\
(I): Suppose a discrete series is collected and we show it by $z_j$
with $j = 1,..., N$.  We partition data with overlapping windows of
length $2n+1$, in such a way that each neighboring segment has $n+1$
overlap points (see Fig. \ref{adap}). Using data in each window of
length $2n+1$, an arbitrary polynomial function, $\mathcal{Y}$, is
constructed. The best polynomial of order $K$ plays corresponding
local trend. To make continuous trend function and to avoid sudden
jump in trend function, we use  following weighted function for
overlap part of $\nu$th segment \cite{hu09}:
\begin{eqnarray}
\mathcal{Y}_{\nu}^{{\rm overlap}}(j)=\left(1-\frac{j-1}{n}\right)\mathcal{Y}_{\nu}(j+n)+\frac{j-1}{n}\mathcal{Y}_{\nu+1}(j)\nonumber\\
\end{eqnarray}
here $j=1,2,...,n+1$. The value of $n$ and the order of fitting function are two free parameters should be determined properly \cite{hu09}.
 In this paper we consider the number of segmentations equal to ${\rm w_{adaptive}}=101$. Also $K=2,4$ and $5$ orders for fitting polynomial are chosen.
The size of each segment is calculated by: $2n+1\equiv2\times{\rm int}\left[\frac{N-1}{{\rm{w}_{adaptive}}+1}\right]+1$. By increasing the value of ${\rm w_{adaptive}}$ and the order of fitting polynomial, almost fluctuations to be disappear, hence the information of the underlying data sets is suppressed.
A schematic of partitioning in the adaptive detrending algorithm has been indicated in Fig. \ref{adap}.
Finally the corresponding adaptive detrended data in non-overlapping segments is given by $x_j=z_j-\mathcal{Y}_{\nu}(j)$ and for overlap part is $x_j=z_j-\mathcal{Y}_{\nu}^{\rm overlap}(j)$.\\
(II): After the first task, clean data produced by adaptive detrended method is used to make profile data as:
\begin{eqnarray}
X(i) &\equiv& \sum_{j=1}^i \left[ x_j - \langle x \rangle \right]
\qquad i=1,\ldots,N
 \label{profile}
\end{eqnarray}

(III): By dividing above profile into $N_s \equiv {\rm int}(N/s)$
non-overlapping segments with equal length, $s$, for each segment
the so-called fluctuation function is computed as follows:
\begin{eqnarray}
&&{\mathcal{F}}(s,m) ={1 \over s} \sum_{i=1}^{s}\{X[(m-1) s+ i] - X_{\rm fit}(i,m)\}^2\nonumber\\
\label{fsdef1}
\end{eqnarray}
for $m=1,...,N_s$
 where $X_{\rm fit}(i,m)$ is arbitrary
fitting polynomial in $m$th segment. Usually, the first order fitting function is considered in above algorithm  \cite{PRL00}.  \\
(IV): The average is defined by:
\begin{equation}
{\mathcal{F}}(s) =\left\{ {1 \over  N_s} \sum_{m=1}^{ N_s} \left [{\mathcal{F}}(s,m)\right] \right\}^{1/2} \label{fdef}
\end{equation}
(V): Finally, the slope of the log-log plot
of ${\mathcal{F}}(s)$ versus $s$ is determined by:
\begin{equation}\label{lam}
{\mathcal{F}}_{x}(s) \sim s^{h}
\end{equation}
For stationary series $H=h$  and for non-stationary data Hurst exponent is $H=h-1$\cite{climate1,sadeghriver,taqqu}. The Hurst exponent
determined by this algorithm will be used to theoretical prediction of $\xi$ defined by Eqs. (\ref{eq8}) and (\ref{eq19}).
In next section all theoretical backbones that clarified up to now, will be applied on well data.

\begin{figure*}[ht]
\begin{center}
\includegraphics[width=1\linewidth]{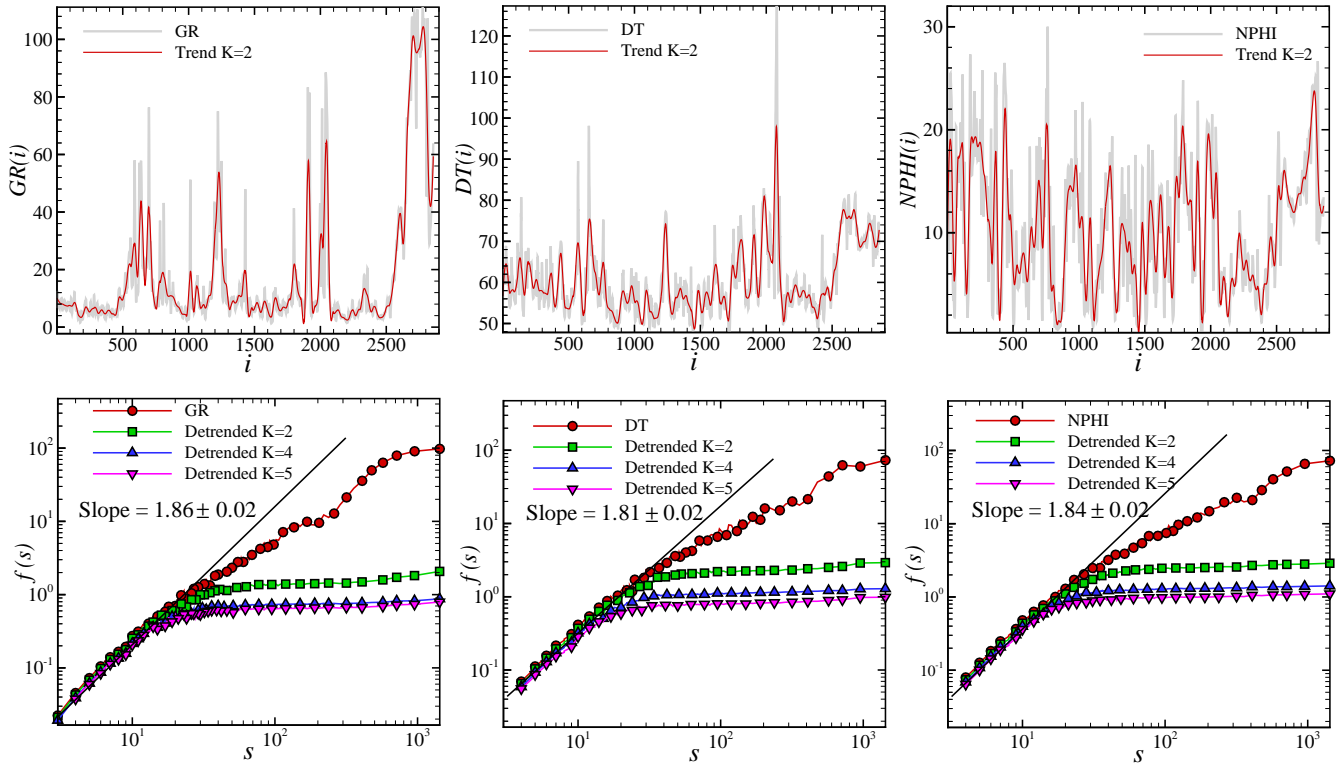}
\caption{\label{fig1} Upper panels show three petrophysical
quantities, Gamma ray (GR), sonic transient time (DT) and Neutron
porosity (NPHI), respectively, versus depth recorded every $15.4$cm
at depth interval $3504.5$m to $3946.8$m for well $\# 2$. In these
panels the gray line corresponds to original fluctuations while the
dark solid line indicates the trend fluctuation constructed by
setting $\rm{w}_{adaptive}=101$ and $K=2$ for fitting polynomial in
each segment. The lower panels illustrate fluctuation function as a
function of scale for different series of well $\# 2$. The filled
circle symbol shows $f(s)$ for original data set. The filled square
symbols are results for clean data by adaptive detrending method
with $K=2$. Up-triangle and down-triangle symbols correspond to
clean data with $K=4$ and $K=5$, respectively.}
\end{center}
\end{figure*}
\section{Data description}\label{data1}

We use well-log data from four oil wells of Maroon reservoir in
southwest of Iran. These data include gamma emission (GR), sonic
transient time (DT) and neutron porosity (NPHI) recorded every
$15.4cm$. The logged interval contains Asmari region formation,
including mainly of fractured carbonate, sand stone, shaly sand and
a trace of anhydrate. Gamma log is a criterion for the natural
radiation of the composition. Gamma emission is received from shales
and shaly sands which have higher radioactivity. Sonic log involves
elapsed time for traveling sound wave through a composition.
Changing the energy of high energy neutrons during their collision
with the component of targets is a benchmark for tracking the
existence of hydrogen in the pore space \cite{Fedi,log1,log2}.
Therefore, NPHI is used for neutron porosity. According to mentioned
criteria, the spatial heterogeneity of properties of the large scale
porous media, such as porosity, density and the lithology at
distinct length scales are determined \cite{log3,log4,log5}. Upper
panels of Fig. \ref{fig1} show GR, DT and NPHI series for well $\#
2$ of this region. In these panels the gray line corresponds to
original fluctuations while the dark solid line indicates the trend
fluctuation constructed by setting $\rm{w}_{adaptive}=101$ and $K=2$
for fitting polynomial in each segment. The lower panels illustrate
fluctuation function as a function of scale for different series of
well $\# 2$. The filled circle symbol shows $f(s)$ for original data
set. The filled square symbols is results for clean data by adaptive
detrending method with $K=2$. Up-triangle and down-triangle symbols
correspond to clean data with $K=4$ and $K=5$, respectively.
Obviously, the $f(s)$ for original fluctuations has not unique
scaling behavior. This situation gives rise for other sets of data
used throughout this paper. Subsequently, one can not determine
associated Hurst exponent, then essentially, adaptive detrending or
every additional method to remove trend in data must be applied in
order to find reliable Hurst exponent. Table \ref{hursttable}
contains the Hurst exponent determined by Adaptive-DFA method. Our
results confirm that all series belong to the nonstationary process
so $H=h-1$. This Hurst exponent is necessary to set up theoretical
prediction represented by Eqs. (\ref{eq8}) and (\ref{eq19}).

\begin{table}
\medskip
\begin{tabular}{|c|c|c|c|c|}
  \hline
    $H$ & $\#1$ & $\#2$ & $\#3$ & $\#4$  \\\hline
  GR & $0.65\pm 0.02$ & $0.86\pm 0.02$ & $0.84\pm 0.02$ & $0.92\pm 0.02$ \\\hline
  NPHI & $0.80\pm 0.02$ & $0.84\pm 0.02$ & $0.76\pm 0.02$ & $0.77\pm 0.02$ \\\hline
  DT & $0.79\pm 0.02$ & $0.81\pm 0.02$ & $0.73\pm 0.02$ & $0.77\pm 0.02$ \\\hline
\end{tabular}
\caption{\label{hursttable} The Hurst exponent, $H$, of data sets recorded
for four wells of the reservoir at $1\sigma$ confidence interval.}
\end{table}

\begin{table}\label{}
\medskip
\begin{tabular}{|c|c|c|c|c|}
  \hline
    $\lambda_0 ^2$ & $\#1$ & $\#2$ & $\#3$ & $\#4$  \\\hline
  GR & $0.042\pm 0.020$ & $0.200\pm 0.002$ & $0.077\pm 0.003$ & $0.040\pm 0.002$ \\\hline
  NPHI & $0.023\pm 0.003$ & $0.045\pm 0.002$ & $0.028\pm 0.002$ & $0.029\pm 0.002$ \\\hline
  DT & $0.002\pm 0.001$ & $0.004\pm 0.001$ & $0.004\pm 0.001$ & $0.005\pm 0.001$ \\\hline
\end{tabular}
\caption{\label{lambdatable} The non-Gaussian parameter, $\lambda_0 ^{2}$ of
data sets recorded for four wells of the reservoir at $1\sigma$
confidence interval.}
\end{table}

\begin{figure*}[ht]
\begin{center}
\includegraphics[width=1\linewidth]{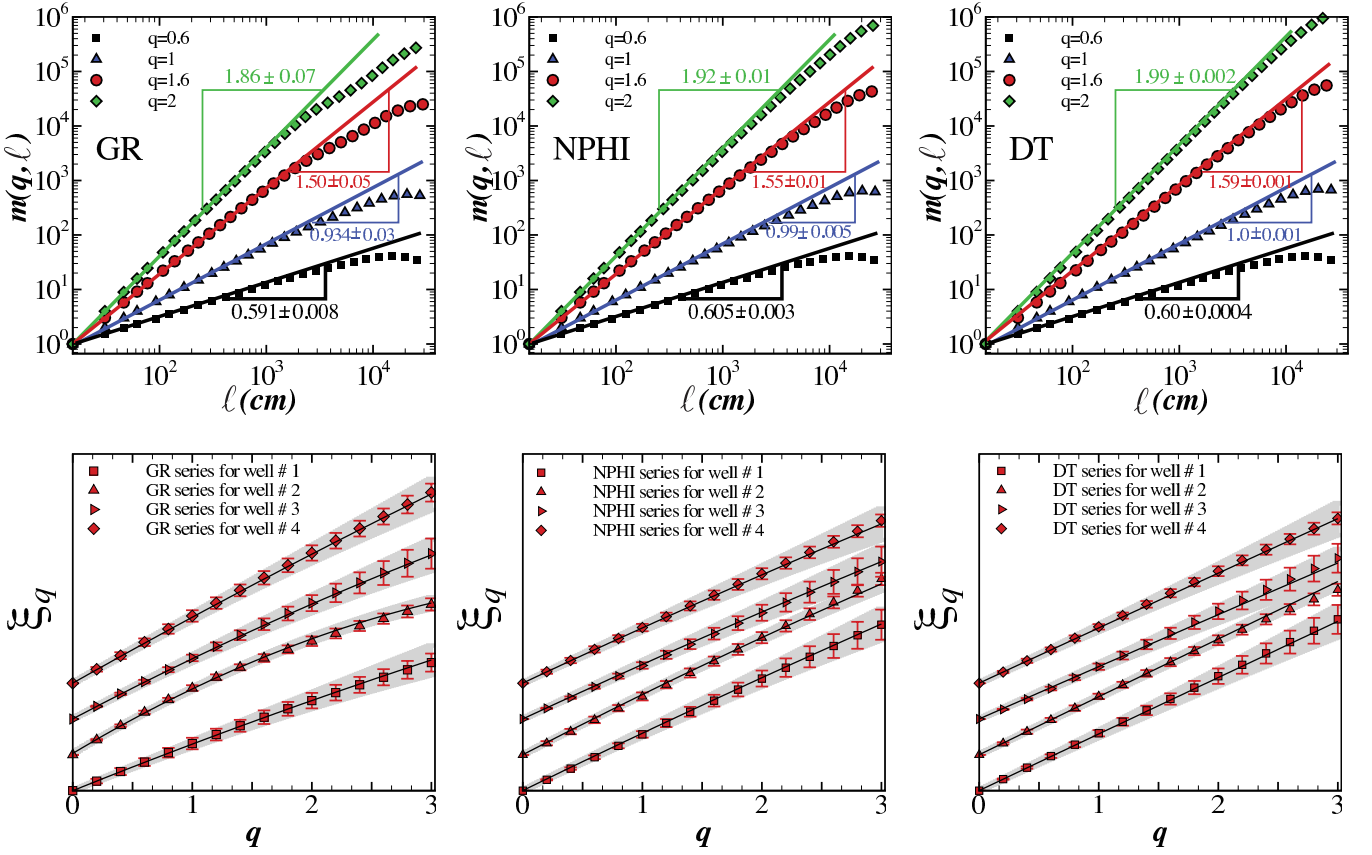}
\caption{\label{fig2} Upper panels show log-log plot of
$m(q,\ell)$ versus $\ell$ for $q=0.6$ (square), $q=1$ (triangle),
$q=1.6$ (circle) and $q=2$ (diamond) for GR (left panel), NPHI
(middle panel) and DT (right panel) for well $\# 2$. The lower
panels correspond to scaling exponent, $\xi_{q}$, versus $q$ from
left to right for GR, NPHI and DT series, respectively for four
wells of the reservoir. Symbols correspond to the empirical series
and solid lines show theoretical prediction given by Eq. (\ref{eq8})
at $1\sigma$ confidence interval corresponding to shaded area.
To make more sense we shifted the value of $\xi_q$ for different data sets vertically.}
\end{center}
\end{figure*}

\begin{figure*}[ht]
\includegraphics[width=0.8\textwidth]{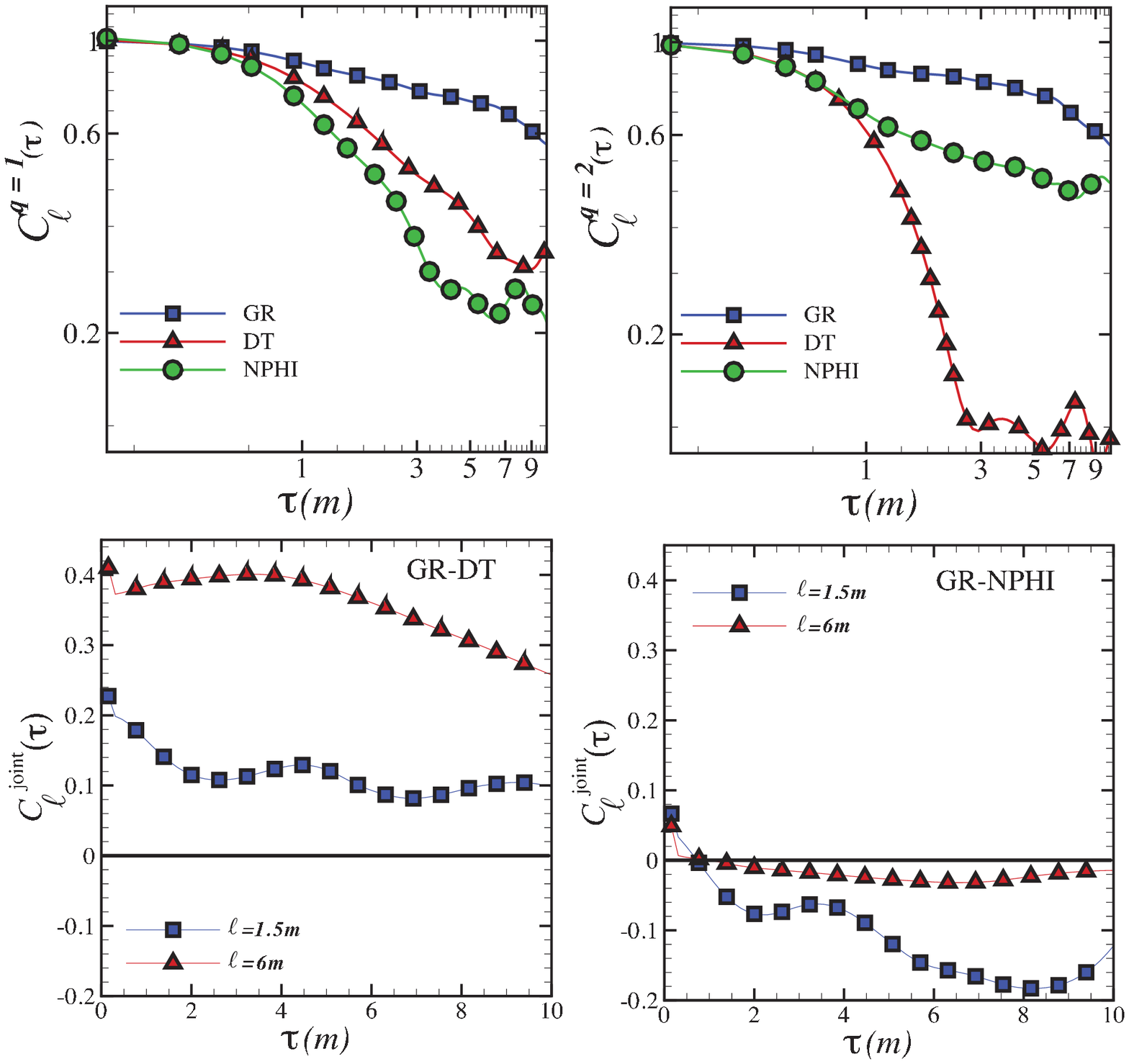}
\caption{Upper panels indicate increment correlation function $C_{\ell}
^q(\tau)$ calculated for $q=1$ (left) and $q=2$ (right) versus lag $\tau$ at scale
$\ell=1.5m$ (Eq.\ref{eq10}). Lower panels illustrate cross-correlation
function, $C_{\ell}^{\rm{joint}}(\tau)$, of the petrophysical series
for well $\# 2$ versus lag $\tau$ at scales $\ell =1.5m$ (square)
and $\ell =6.0m$ (delta) for GR-DT (left) and GR-NPHI
(right).} \label{fig:mainfig}
\end{figure*}

\section{data analysis}\label{data}

In this article we implement multifractal random walk model to
characterize a reservoir by describing some features of
petrophysical quantities. As mentioned before, one reliable method
for multifractal characteristic of a typical data sets is determined
by evaluation of scaling exponent $\xi_{q}$ from the linear state.
To this end, Eq. (\ref{eq1}) is computed for our data. Upper panels
of Fig. \ref{fig2} indicate log-log plot of $m(q,\ell)$ versus
$\ell$ for well $\#2$ at different values of $q$ for various kind of
data sets. These plots verify that $m(q,\ell)$ has scaling nature up
to a typical characteristic scale (this result is satisfied for all
considered wells), consequently, the scaling exponent, $\xi_q$ has
reliable value at $1\sigma$ confidence interval. The scaling
exponent $\xi_{q}$ for the petrophysical quantities has been plotted
in the lower panels of Fig. \ref{fig2}. In this plots symbols
correspond to numerical results. These results demonstrate that
$\xi_{q}$ for all wells is nonlinear for GR and NPHI corresponding
to multifractal nature of mentioned quantities. While $\xi_{q}$ is
almost linear for DT at all considered wells which argues
monofractal behavior. In order to evaluate the theoretical
prediction of $\xi_q$ (Eq. (\ref{eq8})) and to compare it with that
of given by numerical approach, we should determine the
corresponding $\lambda _{0} ^{2}$ and $H$. The multifractal
parameter $\lambda _{0} ^{2}$ is the intercept of $\lambda_{\ell}^2$
versus $\ell$ determined by Eq. (\ref{eq6}) \cite{koohi}. The Hurst
exponent ,$H$, of the data sets has been estimated by adaptive-DFA
\cite{peng94,Hu,bunde02}. The value of Hurst exponent and
multifractal parameter for GR, DT and NPHI have been reported in
Tables. \ref{hursttable} and \ref{lambdatable}. From statistical
point of view, according to Hurst exponent, one can not discriminate
GR, NPHI and DT from each other, while the value of $\lambda_0^2$
can, namely $\lambda_0^2({\rm GR})>\lambda_0^2({\rm
NPHI})\gg\lambda_0^2({\rm DT})$.

Plugging $H$ and $\lambda_0^2$ of each
data in Eq. (\ref{eq8}), theoretical value of $\xi^{\rm The}_q$ is
obtained. Solid lines in the lower panels of Fig. \ref{fig2}
correspond to $\xi^{\rm The}_q$. Our results are consistent with
that of given directly from Eq. (\ref{eq1}) at $1\sigma$ confidence
interval. To make more sense, we shifted the value of $\xi_q$ for different data sets vertically. The multifractal property of GR for four considered wells
expresses that various values of fluctuations in GR series don't
exhibit a global scaling behavior. This means that fractal nature of
Gamma ray series in the reservoir is a local property from the value
of fluctuations point of view. Since DT series is based on the
propagation of sound wave through the media, it follows the
continuum regions as much as possible. Subsequently, the
multifractality would be suppressed. Meanwhile, NPHI data set, has
less multifractality nature than GR series. It represents
inhomogeneity distribution of Hydrogen in the pores.
As mentioned in section \ref{model1}, the larger value of $\lambda_0^2$,
the fatter non-Gaussian tail of PDF resulting in multifractal
behavior.  Namely, in such case, the source of multifractality is
large scale fluctuations corresponding to rare events. Dependency of
multifractal behavior to length scale, $\ell$,  is determined by the
shape of PDF. Koohi et al. \cite{koohi} have shown that for
mentioned data sets of well $\# 2$, non-Gaussianity is scale
invariant for GR and DT, while $\lambda_{\ell}^2$  decreases by
increasing $\ell$. According to current analysis based on $\xi_q$,
one can find a consistency between previous and present results.
Indeed, GR is strongly multifractal and the value of associated
$\lambda_{\ell}^2$ has considerable value. However, the value of
$\lambda_{\ell}^2$ for DT is constant versus $\ell$, but the
individual value is small in comparison to GR's.
Our approach enables us to answer the question concerning the nature
of multifractality. According to the shape of PDF modeled by Eq.
(\ref{eq6}) and the comportment of $\lambda_{\ell}^2$ as a function
of $\ell$ \cite{koohi}, we find that multifractal feature is caused
by the long-range correlations in mentioned series.

In order to investigate the effect of correlations in petrophysical
quantities, the correlation function for $q=1$, and $q=2$, are
calculated by means of Eq. (\ref{eq10}). Upper panels of Fig.
\ref{fig:mainfig} shows the correlation function of the first and
second increment moments of the data sets for well $\# 2$. The
long-range correlation function for GR reveals the global strong
correlations in shaly layers of the reservoir. While, the rapid
suppression of correlation function for second moment of increment
for DT demonstrates that this data set belongs to almost monofractal
category which is consistent with our pervious results regarding
$\xi_q$. In addition, the considerable value of correlation function
for large fluctuations of NPHI demonstrates that this quantity is
non-Gaussian and multifractal at small scales. The same results have
been confirmed for other wells.

\begin{figure*}[ht]
\begin{center}
\includegraphics[width=0.8\linewidth]{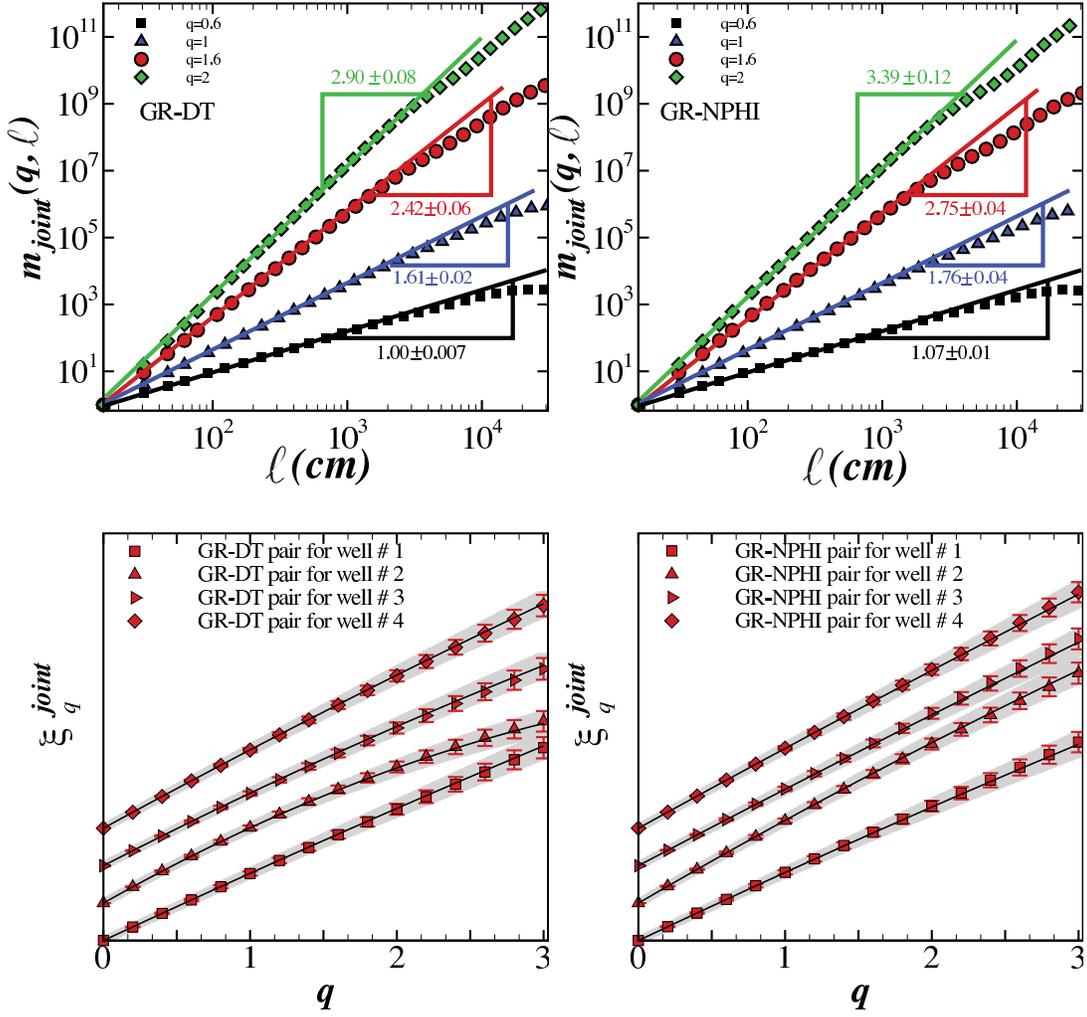}
\caption{\label{fig4} The upper panels show log-log plot of joint
moment $m_{\rm{joint}}(q,\ell)$ with $q_{1}=q_{2}=q$ versus $\ell$
for $q=0.6$ (square), $q=1$ (triangle), $q=1.6$ (circle) and $q=2$
(diamond) for GR-DT pair (upper left) and GR-NPHI pair (upper right) for well $\# 2$. The lower panels indicate joint scaling
exponent, $\xi_q^{\rm{joint}}$, versus $q$ for GR-DT (left) and
GR-NPHI (right) for four wells of reservoir. The symbols correspond
to numerical approach and solid lines represent  theoretical formula
given by Eq. (\ref{eq19}) for $68\%$ confidence interval according
to shaded area. To make more sense, we shifted the value of $\xi_q^{\rm joint}$ for different data sets vertically. }
\end{center}
\end{figure*}

The petrophysical quantities have mutual correlations
in the reservoir. The non-Gaussian PDF of GR at all scales reveals
that GR is playing as a background role in the system that affects
other relevant quantities \cite{koohi}.  Additional investigation in
the context of multiplicative cascades multifractal formalism is
satisfied by cross-correlation of stochastic variances as
\cite{Kiy06}:
\begin{equation}
C_{\ell}^{\rm joint} (\tau) \equiv\left \langle  \left[\bar{\omega}^{(\ell)}_{1} (i+\tau)-  \langle \bar{\omega}^{(\ell)}_{1}\rangle    \right]\left[ \bar{\omega}^{(\ell)}_{2}(i)-\langle  \bar{\omega}^{(\ell)}_{2}\rangle     \right] \right\rangle
\end{equation}
with
\begin{eqnarray}
\bar{\omega}^{(\ell)}_{\diamond} (i)&=& \frac{1}{2}\ln \sigma^2_{\diamond}(\ell;i)\\
\sigma^2_{\diamond}(\ell;i)&=&\frac{1}{\ell}\sum_{j=1+(i-1)\ell}^{i\ell} \delta_{\ell}x^2_{\diamond}(j)
\end{eqnarray}
where $(\diamond)$ is replaced by $(1)$ and $(2)$ for first and
second data sets in the underlying pair. To explore the nature of
cross-correlation for  series we compute $C_{\ell}^{\rm joint}
(\tau)$ for GR-NPHI and GR-DT pairs associated to well $\# 2$. Fig.
\ref{fig:mainfig} indicate the cross-correlations at scale $\ell$ as
a function of $\tau$. For GR-DT pair, cross-correlation has positive
sign and behaves as long-range phenomenon for $\ell=1.5$m and
$\ell=6.0$m. For GR-NPHI, there is an anti-correlated behavior and
by increasing $\ell$, the magnitude of mutual interaction
asymptotically goes to zero. Positive cross-correlation for GR-DT
probably corresponds to existence of continuum shaly region in which
sonic sound prefers to pass it, therefore fluctuations in this pair
are synchronized resulting in possessing positive cross-correlation
based on stochastic variance. However, the negative
cross-correlation for GR-NPHI pair implies the fact that regions
with shaly layers prevent gathering of Hydrocarbon or water in the
pores. To make more sense and to quantify the joint multifractality,
we rely on approach explained in section \ref{model2}, and determine
the value of joint multifractal parameter, $L_0$ \cite{Muzy2}. The
values of $L_0$ for GR-DT and GR-NPHI at $1\sigma$ confidence
interval for four considered wells in the reservoir are reported in
Table \ref{Tb3}. These values demonstrate that the nature of joint
multifractality is related to magnitude of cross-correlation and
they are compatible with previous interpretations. Using Eq.
(\ref{xijoint1}), the slope of log-log plot of $m_{\rm
joint}(q_1,q_2;\ell)$ versus
 $\ell$ for $q_1=q_2$ gives $\xi_q^{\rm joint}$. Upper panels of Fig.
 \ref{fig4} illustrate the log-log plot of $m_{\rm joint}(q,\ell)$ versus $\ell$ for pairs of GR-DT and GR NPHI for
well $\# 2$. Lower panels of Fig. \ref{fig4} correspond to the
numerical and theoretical values of $\xi_q^{\rm joint}$ for mentioned
pairs and for four wells of the reservoir.
In order to estimate the theoretical prediction of  $\xi_q^{\rm
joint}$, we use Eq. (\ref{eq19}) and take into account the relevant
values of $\xi_q$ and $L_0$. The solid lines in the lower
panels of Fig. \ref{fig4} correspond to this approach and shaded
area indicates the $68\%$ confidence interval. Joint multifractality is positive for
GR-DT causing the convex shape for $\xi_q^{\rm joint}$, while the
negative value of $L_0$ reduces this convexity in
$\xi_q^{\rm joint}$ for GR-NPHI.


\begin{table*}[ht]
\medskip
\begin{tabular}{|c|c|c|c|c|}
  \hline
    $L_0$ & $\#1$ & $\#2$ & $\#3$ & $\#4$  \\\hline
  GR-NPHI & $-0.004\pm 0.010$ & $-0.068\pm 0.002$ & $-0.032\pm 0.001$ & $-0.015\pm 0.001$ \\\hline
  GR-DT & $0.003\pm 0.010$ & $0.062\pm 0.001$ & $0.021\pm 0.001$ & $0.016\pm 0.002$ \\\hline

\end{tabular}
\caption{\label{Tb3} The joint multifractal parameter, $L_0$
of data sets for four wells of the reservoir at $1\sigma$ confidence
interval.}
\end{table*}

\section{Summary and Conclusion}\label{sum}
In this paper, we relied on multifractal random walk and joint-multifractal random walk approaches to analyze
some features of petrophysical quantities represented by GR, NPHI
and DT as some of petroleum reservoir indicators. Mentioned data
sets have been collected in well-logging through four wells in
Maroon reservoir in southwest of Iran.

To infer statistical information through statistical indicators, one
must take care about following strategy:  when there are more than a
few indicators existing in a typical phenomenon, it is important to
estimate how efficient they are and how they are cross-correlated. To this end, we should determine
the degree of correlation between mentioned indicators.
In other words, if the indicators are cross-correlated to each
other, probably, the content of their information is less than two
completely independent indicators. To analyze oil wells, lots of
indicators have been introduced and without having knowledge about
their cross-correlations,  results coming from each indicators are
not reliable. In the oil wells, as we get closer to oil reservoir,
the properties of the medium changes. The effect of this variation
causes a  non-Gaussian  behavior of indicators, hence joint
multifractal random walk could be a useful measure to examine this
kind of cross-correlation between these indicators. multifractal
random walk model has been introduced according to the concept of
multiplicative random cascades in multifractal
formalism\cite{Bacry,Muzy1}. The parameter which controls the
strength of multifractality is $\lambda_0^{2}$. The larger value of
$\lambda_0^{2}$, the higher probability of finding large scale
fluctuations in a system causing a non-Gaussian PDF and strong
multifractality. This gives rise to non-linear scaling exponent of
absolute moment of fluctuations, $\xi_q$, (Eq.  (\ref{eq1})) versus
$q$. In order to explore the mutual effects, joint multifractal
random walk of data sets have been considered under the notion of
joint multiplicative cascade processes. The cross-correlation in the
fluctuations of stochastic variances causes the joint
multifractality represented by joint multifractal parameter, $L_0$.
According to the multiplicative approach, there is a relation
between $\xi_q$ and $\lambda_0^2$ which allows us to check the
consistency of both approaches (Eq.  (\ref{eq8})). In addition, for
joint analysis, theoretical prediction for $\xi_q^{\rm joint}$ using
the value of $H$, $\lambda_0^2$ and $L_0$ exponents can be written
according to Eq. (\ref{eq19}). The positive or negative value of
$L_0$ is due to the existence of persistent or anti-persistent
correlation of large fluctuations of two underlying series. The
positive value of $L_0$ causes the convex shape of $\xi^{\rm
joint}_q$, however, the negative value of $L_0$ decreases the
convexity of $\xi^{\rm joint}_q$.

Our results demonstrated that GR in Maroon reservoir exhibits strong
multifractality due to the almost high value of $\lambda_0^2$,
consequently $\xi_q$ is non-linear (see Fig. \ref{fig2}). Also its
PDF is non-Gaussian and scale-invariant \cite{koohi}. This
demonstrates high probability of the occurrence of large
fluctuations in GR series which emitted from shaly layers and leads
to multifractal property. While the value of $\lambda_0^2$ for DT
can be ignored (Table \ref{Tb3}). This gives rise to a linear function for $\xi_q$
which is a hallmark of monofractal behavior. This phenomenon is
explained based on the fact that sound wave follows the continuum
regions in the reservoir, hence this quantity exhibits monofractal
property. For NPHI data set, since the shape of PDF is scale
dependent \cite{koohi} and its $\lambda_0^2$ is less than GR's, our
analysis indicated  $\xi_q$ is non-linear. Indeed, inhomogeneity
distribution of Hydrogen in the pores at small scales results in
multifractal behavior of NPHI only at small scales.

Joint analysis of data sets for all mentioned wells of Maroon
reservoir proved the negative and positive values of $L_0$ for
GR-NPHI and GR-DT pairs, respectively. Computational value of joint
moments, $\xi_q^{\rm joint}$ displayed  that joint multifractality
for GR-DT is almost larger than GR-NPHI pair (Fig. \ref{fig4}).
Lower panels of Fig. \ref{fig4} proved that the numerical and
theoretical prediction for $\xi_q^{\rm joint}$ are in agreement at
$1\sigma$ confidence interval. From petrophysical point of view, one
can mention that sonic sound prefers to pass through the continuum
shaly region, so we expect the positive cross-correlation between
GR-DT pair corresponding to statistical synchronization of
fluctuations of mentioned pair, $L_0>0$. In addition, in the
presence of shaly layers the probability of finding pores in media
decreases, subsequently we expect negative cross correlation between
GR and NPHI indicators corresponding to $L_0<0$. Finally, it could
be interesting to apply these methods to asses other indicators and
available data of other reservoirs to examine other effects.

\textbf{Acknowledgments:} Authors are grateful to H. Dashtian for
preparing data used in this research. S.M.S.M. is tankful to Associate office of ICTP. 

\section{APPENDIX}

In this appendix by relying on multivariate form of PDF, we show  a
proof in details for the relations, given in Eqs.(\ref{eq19}) and
(\ref{eq20}). Suppose an stochastic increment for a bivariate
process $\textbf{x}(t)$ with lag $\ell$ as:
\begin{eqnarray}
\delta_{\ell}\mathbf{x}(t)&\equiv&\delta_{\ell}x_1(t)\delta_{\ell}x_2(t)\nonumber\\
&=&\left[x_1(t+\ell)-x_1(t)\right]\left[x_2(t+\ell)-x_2(t)\right]
\end{eqnarray}
For convenience, we denote the processes $x_1(t)$ and
$x_2(t)$ by $x_1$ and $x_2$, respectively. The joint moment of the processes reads as:
\begin{eqnarray}\label{eq23}
&&m(q_1,q_2,\ell)\equiv\langle|\delta_\ell x_1|^{q_1}|\delta_\ell x_2|^{q_2}\rangle\nonumber\\
&=& \int
|\delta_\ell x_1|^{q_1}|\delta_\ell x_2|^{q_2}P_\ell\left(\delta_{\ell}
x_1,\delta_{\ell} x_2\right)d(\delta_{\ell} x_1)d(\delta_{\ell} x_2)\nonumber\\
\end{eqnarray}
where $P_\ell\left(\delta_{\ell} x_1,\delta_{\ell} x_2\right)$ is
equivalent to Eq. (\ref{eq16}) and can be defined as follows:
\begin{eqnarray}\label{eq24}
P_{\ell} (\delta_{\ell} x_1,\delta_{\ell} x_2)&=&\int\int G_{\ell,\beta \times
\ell} (u_1,u_2) e^{-(u_1+u_2)} \nonumber\\ &&P_{\beta \times \ell} (e^{-u_1}
\delta_{\ell} x_1,e^{-u_1} \delta_{\ell} x_2) du_1 du_2\nonumber\\
\end{eqnarray}
$\delta_\ell x_1$ and $\delta_\ell x_2$ have scaling behavior:
\begin{equation}\label{eq25}
\delta_\ell x_1=\ell^{H_1}\delta x_1 \ \ \ \ ,\ \ \ \
\delta_\ell x_2=\ell^{H_2}\delta x_2
\end{equation}
Plugging Eqs.(\ref{eq24}) and (\ref{eq25}) in Eq. (\ref{eq23})
then, by changing the variables:
\begin{equation}
 x'_1\equiv e^{-u_1} \delta x_1 \ \ \ \ , \ \ \ \
 x'_2\equiv e^{-u_1}\delta x_2
\end{equation}
the joint moment, Eq. (\ref{eq23}), becomes:
\begin{widetext}
\begin{equation}\label{eq27}
m(q_1,q_2,\ell)=\ell^{q_1H_1}\ell^{q_2H_2}\left(\int\int(x'_1)^{q_1}(x'_2)^{q_2}P(x'_1,x'_2)d
x'_1 dx'_2\right)\left(\int\int e^{q_1u_1}e^{q_2u_2}G_{\ell}
(u_1,u_2)du_1 du_2 \right)
\end{equation}
\end{widetext}
The first double integral in the above equation introduces the
prefactor:
\begin{equation}
\mathcal{A}_{q_1,q_2}=\int\int(x'_1)^{q_1}(x'_2)^{q_2}P(x'_1,x'_2)d x'_1
dx'_2
\end{equation}
where $P(x'_1,x'_2)$ is a joint Gaussian PDF:
\begin{equation}
P(x'_1,x'_2)=\frac{1}{2\pi\sqrt{\rm{Det}(\mathbf{\Sigma}_{(\ell)})}}\
\exp\left({-\frac{\mathbf{x'}^T
\cdot\mathbf{\Sigma}_{(\ell)}^{-1}\cdot\mathbf{x'}}{2}}\right)
\end{equation}
with covariance matrix, $\mathbf{\Sigma}_{(\ell)}$ represented by:
\begin{equation}
\mathbf{\Sigma}_{(\ell)}\equiv\left(
               \begin{array}{cc}
                 \Sigma_{(\ell)}^{11}& \Sigma_{(\ell)}^{12}\\
                 \Sigma_{(\ell)}^{21} & \Sigma_{(\ell)}^{22} \\
               \end{array}
             \right)
\end{equation}

  The
prefactor $\mathcal{A}_{q_1,q_2}$ can be written as two separated integrals:
\begin{eqnarray}
\mathcal{A}_{q_1,q_2}&=&\frac{1}{2\pi \sqrt{(1-S_\ell^2)}}\int
(x'_1)^{q_1} e^{-\frac{(x'_{1})^{2}}{2(1-S _\ell^2)}} dx'_{1}\nonumber\\ &&\int (x'_2)^{q_2} e^{-\frac{(x'_{2})^{2}}{2(1-S_\ell^2)}}dx'_{2}
\end{eqnarray}
where covariance coefficient, $S_\ell$, is defined as
$S_\ell=\frac{\Sigma_{(\ell)}^{12}}{\sigma_1(\ell) \sigma_2(\ell)}$. In order to
calculate the second integral in Eq. (\ref{eq27}), Fourier transformation
of $G_{\ell}(u_1,u_2)$ is implemented which is defined as:
\begin{equation}
G_{\ell}(k_1,k_2)=e^{\ln \ell(i\mathbf{k}^T\Gamma-\frac{1}{2}\mathbf{k}^T.\Lambda. \mathbf{k})}
\end{equation}
where $\Gamma$ is bivariate mean vector, $\Gamma=(\Gamma_1,\Gamma_2)$. Thus,
the second integral determining the scaling dependence of the joint
moment is estimated as:
\begin{widetext}
\begin{equation}\label{eq32}
\int\int e^{q_1u_1}e^{q_2u_2}G_{\ell}(u_1,u_2)du_1 du_2=
{\ell}^{q_1\Gamma_1+q_2\Gamma_2-\frac{1}{2}\left(\lambda_1^2q_1^2+\lambda_2^2q_2^2\right)-q_1q_2L_0}
\end{equation}
\end{widetext}
In order to obtain the mean values of $\Gamma_1$ and $\Gamma_2$, we
assume the compact support case for which in absence of cross
correlation ($L_0=0$) the scaling exponent of multifractal
portion vanishes for $q_1=q_2=1$. This yields
$\Gamma_1=\frac{\lambda_1^2}{2}$ and $\Gamma_2=\frac{\lambda_2^2}{2}$,
thus the joint moment becomes:
\begin{equation}
m(q_1,q_2,\ell)=\mathcal{A}_{q_1,q_2}\ell^{\xi_{q_1}^{(1)}+\xi_{q_2}^{(2)}-q_1q_2L_0}
\end{equation}
where $\xi_{q_i}^{(i)}=q_iH_i-\frac{\lambda_i^2}{2}q_i(q_i-1)$
with $i=1,2$ is the scaling exponent of each single process.



\end{document}